\documentclass[12pt]{article}		
\usepackage{fancyhdr}
\usepackage{multicol}
\usepackage{latexsym}
\usepackage{graphicx}
\usepackage[T1]{fontenc}			
\usepackage{setspace}				
\usepackage{marvosym}				
\usepackage{seqsplit}				
\usepackage[hyphens]{url}

\usepackage[round,semicolon,authoryear]{natbib}	
\begin{document}

\title{Practical Judgment, Virtue, and Intuition in the Use of Opaque AI-Enabled Systems}	
\author{Nathan Gabriel Wood | nathan.wood\MVAt tuhh.de
	\\Andrew P. Rebera | rebera.andrew\MVAt gmail.com}
\date{}

\maketitle	


\begin{abstract}

AI-enabled systems are seeing increasing deployment across numerous domains, with many being ``black boxes'' with respect to core functions and capabilities. I.e., many systems take inputs and give outputs, but without users having any ability to see how the former lead to the latter. AI-enabled systems are also being used to augment autonomy in systems, and autonomy coupled with opacity raises numerous concerns surrounding, e.g., the reliability of systems, their regularity in functioning, human ability to control them, or whether deploying opaque and potentially autonomous systems is in compliance with ethical and legal norms. In this article, we argue that many of these worries can be mitigated by leveraging practical judgment, virtue, and intuition in the deployment and use of opaque AI-enabled systems. We show that focusing on these distinctly human capabilities provides a means for bridging between the practical challenges created by opacity and the ethical, legal, and social norms underpinning particular domains. We argue that a core element in doing this is a recognition that many positive human traits are not quantifiable and we therefore must develop training regimen and guidelines on AI deployment anchored in humanistic but non-quantifiable values. Throughout the article, we focus on the military domain as an exemplar of the importance of practical judgment, virtue, and intuition as drivers for ethical and effective human decision-making surrounding AI deployments, but the underlying arguments apply to all domains where opaque and potentially autonomous systems are being deployed (subject to domain-specific alterations).

\bigskip
	
\noindent\textbf{Keywords:} \emph{AI, Artificial Intelligence, Opacity, Autonomous Systems, Ethics, Virtues} 

\end{abstract}


\section{Introduction}\label{sec_intro} 

Across many domains, AI-enabled systems are complementing and in some cases outsourcing human capabilities. Up until 2022, most deployments of AI were in the form of narrow systems designed and built using symbolic AI techniques and/or machine learning (ML) and tailored datasets to create task-specific systems \citep{babu2024study,wu2025ai,markose2026narrow}. However, with the development of generative pre-trained transformers (GPTs) and the highly successful launch of ChatGPT in November of 2022, there has been an explosion of interest in more generally applicable AI systems, in particular LLMs, for a variety of tasks. However, any ML system trained on large datasets and using deep neural networks (DNNs) is apt to be opaque, meaning there is no reliable method for users or even developers to see how a system goes from particular inputs or prompts and arrives at discrete outputs \citep{loyola2019black,fomin2023black}. This raises challenges from ethical, legal, and operational standpoints, as, among other things, the behavior of opaque systems cannot be as reliably predicted and it is more difficult (perhaps impossible) to rule out certain behaviors or outputs. Opacity can also undermine effective human-machine teaming, as systems whose functioning is unknown to humans (and possibly unknowable) may undermine how well humans can make use of such systems. 

In response to the challenges posed by opaque systems, many solutions or mitigation strategies have been proposed. Some focus on institutional approaches \citep{alizadeh2025unveiling}, others on education of users or the implementation of rough guidelines to avoid the most dangerous or unwanted uses \citep{barman2024beyond}, but one of the most ambitious responses -- and certainly the most technologically demanding -- is to develop so-called explainable AI systems (XAI) which (partially) remove opacity by having systems provide explanations of their behavior and processes \citep{gunning2019darpa,gunning2019xai,gunning2021darpa,das2020opportunities,speith2022review,kalasampath2025review}. XAI sees particularly strong promotion in domains where opaque AI-enabled systems might be used for safety-critical tasks, such as in medicine \citep{tjoa2020survey,sheu2022survey} or the military \citep{christie2023regulating,kwik2023performance,lindelauf2025building}.\footnote{Cf. \citep{wood2024explainable} for arguments critical of XAI in the military domain.} This is especially the case when such AI-enabled systems are embedded in platforms with (some) autonomous functionality, as coupling opacity with autonomy exacerbates basic challenges posed by opacity and introduces additional concerns surrounding, e.g., responsibility attribution,\footnote{On ``responsibility gaps'', see \citep{matthias2004responsibility,asaro2006should} (general discussion), \citep{sparrow2007killer,sharkey2007automated,pagallo2011robots} (in the military), \citep{oimann2023responsibility} (overview of the military debate). Cf. \citep{hindriks2023risks,wood2023awsresponsibility,rebera2024reactive,kiener2025responsibility} (responses and further developments).\label{ftnote_responsibility-gap}} deskilling \citep{ferdman2025deskillingreason,ferdman2025deskillingstructural,natali2025deskilling}, or retaining meaningful human control over systems \citep{article362013killer,hrw2018heed,amoroso2020autonomousmhc,schwarz2021autonomous}, showcasing a serious need for solutions. Focusing on technical and human limitations and seeking to address these through technological solutions ignores, however, the ways distinctly human traits can be leveraged to mitigate many of the challenges opacity and autonomy bring. Indeed, some basic characteristics of humans which make us liable to make certain (kinds of) mistakes when deploying autonomous and opaque systems can actually be capitalized on to ground positive proactive approaches to responsibly and effectively deploying emerging technologies. 

In this article, we argue that practical judgment, virtue, and intuition can be crucial tools for mitigating numerous challenges posed by opaque AI-enabled systems, especially when such systems are embedded in autonomous platforms. We show that by starting from a virtue ethics approach which highlights individual character and the centrality of practical judgment, systems which are not fully understood (or potentially even understandable) for certain users may still be responsibly and effectively used, so long as humans deploying the systems have well-developed characters and judgment and minimal sufficient knowledge of systems' functionality. We also draw out a critical and often overlooked aspect of judgment, namely that it may be treated as a (finite) resource, showing how this understanding can help underpin responsible and context-sensitive uses of opaque systems.

Before moving onto the arguments though, two preliminary points are worth addressing. First, many things are labeled as ``AI'', with hype around this class of technologies muddying the intellectual space greatly \citep{tucker2022artifice,marx2023fall,floridi2024why,mattmann2024ai}. Moreover, given how many types of systems genuinely do incorporate AI, and for such a variety of tasks, there are good grounds for us to stop simply saying ``AI'' altogether and instead provide clarity on what exact systems we have in mind \citep{wood2026stop}. In what follows, our primary concern will be systems which make use of machine learning and large datasets or which rely on techniques which lead systems to be opaque. Thus, symbolic AI and certain small ML trained models fall outside the scope of our presentation, as symbolic AI systems are generally transparent (\citep[p. 34]{vanengers2019governmental}; \citep{toy2023transparency}), and small ML models may be transparent due to their limited size and complexity. However, once models become sufficiently large, it becomes practically impossible to understand their inner workings, and some AI techniques guarantee opacity. It is these which are our concern.\footnote{We will not provide a definition of ``AI'' here, but for classic and more contemporary views, see, e.g., \citep[p. 116]{newell1976computer}; \citep[p. 308]{mccarthy1988mathematical}; \citep[p. 19]{wang2019defining}; or \citep[p. 22]{russell2021artificial}.}

Second, our arguments apply to the use of opaque AI-enabled systems generally, especially such systems when they have autonomous functionality. However, for concreteness in presentation, we will focus on AI use in the military, as virtue ethical approaches are already common in military institutions, practical judgment is a core competency soldiers are trained for, and the challenges of opacity and autonomy in systems are highly pronounced in military contexts. Nonetheless, there are many domains which feature some or all of these points, with medicine showing significant overlap, and the core arguments and conclusions will apply to the use of opaque systems more generally (subject to domain-specific modifications). 

\section{Opacity and Its Challenges}\label{sec_opacity} 

\subsection{Opacity}\label{subsec_opacity_def.-opacity}

Above, opacity was described in terms of users or developers not being able to see how systems arrive at outputs from given inputs, thus rendering them ``black boxes''. More precisely, opacity ``is best understood to define the degree of unknown elements in a machine'' and ``we can understand an `opaque system' to be one that is simply not known in further detail'' \citep[p. 30]{kempt2024unexplainable}. Following \citep{burrell2016how}, a central and often underappreciated feature of opacity is that it can be tied to both systems and users.\footnote{Explanations are likewise in crucial respects user- and task-dependent. See \citep{asghari2021explain,langer2021what}.} Thus, certain systems may be \textit{in principle} understandable and transparent, but \textit{in practice} opaque for users lacking capacities necessary for successfully interrogating the systems. Jenna Burrell identifies one part of this form of opacity as ``Opacity as technical illiteracy''\citep[p. 4]{burrell2016how}, but practical human limits need not be viewed only through the lens of ``illiteracy''. Indeed, some humans competent to \textit{in principle} understand a machine's functioning may \textit{in practice} be unable to because the sheer number of interconnections, nodes, or weights in a model make it infeasible for a human to interrogate the full system. 
This problem is further exacerbated by the rate at which emerging technologies are being iterated upon, as ML systems which are retrained in the course of months or weeks may not allow for full reviews of models before they are made obsolete by subsequent versions.

Systems may also be in principle opaque, meaning no human, either user or developer, may be in a position to understand it. This can be due to a number of reasons: the underlying mathematics or natural laws governing systems may not (yet) be known; systems may have multiple interconnected layers whose meaning or connection to tasks are not fully identifiable for humans; or systems may have authority to rewrite their programs during operation, making it impossible for humans to see with certainty, after deployments, what code was actually running during operation.

For ML-based systems, especially those relying on large datasets for training, there will be at minimum a very high -- and potentially insurmountable -- degree of \textit{in practice} opacity due to how large training data sets are, how many nodes and weights are present in models, and how many interconnections there are between elements of models. Additionally, because machine learning involves systems themselves forming models of patterns in data and establishing weights between data points, there will often be \textit{in principle} opacity as well. This is because such systems have self-determined a portion of their programs without a human necessarily overseeing the process, and without a human necessarily even knowing what elements of the resulting programs are due to direct human inputs, which are logical implications of human inputs, and which are statistically learned from training data. In any event, virtually all ML systems trained on large datasets will be both \textit{in practice} and \textit{in principle} opaque with regards to core aspects of their training process and functionality. 

\subsection{Basic Challenges of Opacity}\label{subsec_opacity_challenges}

Unfortunately, while numerous narrow applications of AI are deterministic or admit only minor opacity, 
the types of AI increasingly in use in society at large and in various critical domains are in practice and in principle opaque for all of the above reasons. For example, large language models (LLMs), generative AI (GenAI) systems used for audio, image, or video generation, or any system based on neural networks 
will lack explainability and be both in practice and in principle opaque \citep{raleigh2025clarifying,sogaard2023opacity}. This opacity can make systems unpredictable for users, lower users' trust in systems, make usage of systems potentially negligent (if unpredictability is apt to lead to dangerous outcomes users can foresee), and it may undermine how well or meaningfully humans can control systems.\footnote{For these various concerns, see, e.g., \citep{fraser2022ai,davidovic2023purpose,robbins2023many,sogaard2023opacity,toy2023transparency,freiman2025opacity,raleigh2025clarifying,boisseau2026expertise}.} Much of the research on challenges of opacity focuses on ethical and legal issues attending use of such systems, but it is also critical to bear in mind that all of these worries present operational risks as well, 
as unpredictability, lowered trust, potential legal negligence, and loss of broad societal acceptance of systems may all negatively impact how effectively they can be used. 
In short, opacity can be the source of significant challenges arising from numerous normative and organizational sources. 

\subsection{Opacity and Autonomy}\label{subsec_opacity_opaque-autonomous}

Coupling machine learning with autonomy can exacerbate all of the concerns sketched above. Very briefly, autonomous systems may be conceived of as systems which can execute core tasks or ``critical functions'' normally requiring human judgment \citep[p. 5]{icrc2014autonomous}, and do so without contemporaneous human input or oversight.\footnote{Such definitions can be found for autonomous systems in, e.g., the military (\citep{williams2015defining,boothby2016weapons,caron2020defining}; \citep[p. 1]{icrc2021positiona}; \citep[p. 21]{usdod2023directive3000.09}; \citep{wood2023awsclarification,pacholska2024autonomous}), medicine (\citep{moustris2011evolution}; \citep[p. 1]{yang2017medical}; \citep{haidegger2019autonomy}), and for machine/AI systems generally \citep{fahimi2009autonomous,ezenkwu2019machine,walsh2021autonomy}. Note that some debates discuss machines ``making decisions'' or similar concepts (e.g., \citep{heyns2016autonomous}), but from philosophical, ethical, legal, and technical standpoints, machines cannot be said to ``decide'', ``choose'', ``act'', or otherwise engage in agential activities. Rather, they ``execute processes'' in line with their programming or construction (see \citep{wood2026extensions}).} 

When systems are opaque and capable of carrying out such ``critical functions'' autonomously, this compounds the risks involved. During the use of a merely opaque but still fully human-controlled system, there is usually a persistent ability for the system to be constrained or, if need be, turned off. 
This means if one of the risks from opacity does materialize, say, the system being potentially unpredictable, then there is still an opportunity for the human overseeing the system to mitigate if not avoid unwanted consequences of that unpredictability. 

For concreteness, suppose security services are using an opaque ML-trained object identification system to flag potential threats, and due to that opacity, the system might give recommendations or analysis which users could not predict. Assume it flags some unarmed individual as a threat. Nothing beyond this happens though, as the system is not autonomous and a human must decide on a next course of action. On reviewing video footage, the human sees the flagged individual is not carrying a weapon, and the system must have, due to some issues in its training set or learned patterns, mistakenly given that assessment, and done so in a way that might not have been predictable to users. Because a user was responsible for what would happen next though, the error which was unforeseen due to system opacity could be prevented from leading to further tragedy. 

If that same opaque object identification system is embedded in an autonomous weapon system which can select and engage targets identified as threats, it is clear how quickly problems may arise. Opacity in the former system may prevent users from knowing what might lead the system to make mistakes in identifying objects, making it potentially impossible to add deployment parameters or limit deployments to address possible mistakes -- because users may not even know what mistakes could occur, or why. Giving the opaque system authority over autonomous engagement could therefore lead to unforeseen and potentially unforeseeable mistaken engagements. These engagements may also be fully outside of human control -- say, because the system is operating beyond the limits of operator connection range -- meaning there may be no one who could halt a mistaken engagement once it has begun. 

Similar problems arise with regards to all of the other challenges of opacity discussed. Embedding opaque systems within autonomous systems leads to deep issues for control, predictability, and both responsible and effective deployment of emerging technologies. And while some of these challenges may be partially addressed through technical or institutional fixes, 
it is critical we also recognize how basic humanistic approaches may be leveraged to improve the efficacy and robustness of efforts to address these challenges. 

\section{Virtue, Practical Judgment, and Intuition}\label{sec_virtue-judgment-intuition} 

Opaque and autonomous systems do not exist in a vacuum, but participate in broader contexts, alongside other actors and decision-makers (some of whom are human and some of which may not be). The broader socio-technological systems in which opaque systems are embedded can be designed and maintained to promote effective human involvement and accountability.

Human involvement is valuable because humans have capabilities machines do not, for example the capacity for moral responsibility. Including humans prominently in the setups in which opaque systems are embedded ensures appropriate chains of responsibility are maintained or established.\footnote{On ``responsibility gaps'', see note \ref{ftnote_responsibility-gap}.} Well-known strategies for this include human-in-the-loop (HITL) and human-on-the-loop (HOTL) approaches \citep{hrw2012losing,sipri2017mapping,calvert2024designing}. With HITL, systems require human intervention to complete their task (e.g., a human must approve procedure from one (sub-)decision or process to the next). With HOTL, systems may execute subsequent processes freely, but under the supervision of a human who can intervene.

The difficulty with these approaches is that humans being in- or on-the-loop does not necessarily entail they have the ability to intervene as required -- it does not ensure meaningful human control \citep{article362013killer,article362013structuring,roff2016meaningful,santonidesio2018meaningful};\footnote{Cf. \citep{wood2026extensions}.} humans may have sub-optimal understanding of systems or their impacts; they may not be fully able to monitor or intervene in all relevant steps in decision-processes; they may not be physically capable of intervening (e.g., no human is fast enough to intervene in high-frequency trading) \citep{mackenzie2021trading}; and they may not be cognitively or psychologically capable of intervening (e.g., if affected by automation bias \citep{johnson2022ai}).

One response to such concerns is to task oversight to another machine, a machine unaffected by fatigue, distraction, cognitive overload, etc. In some cases, this may be appropriate. Yet humans bring added value beyond the capacity for taking responsibility (or being held responsible) for outcomes, as humans have oversight capacities machines lack: humans are capable of appreciating, understanding, and responding to situations in ways machines cannot, capacities which are uncodifiable. (If they were codifiable, they could -- and likely would -- be coded into the opaque system or the supervisory systems overseeing it.) Thus, these human abilities are not reducible to the application of a rule-book, but rather involve a more intuitive sensibility and capacity for practical judgment.

The idea of a specific, perhaps even \textit{sui generis}, sensitivity towards salient features of situations, and the capacity to respond on the basis of it, is common to discussions of expertise or excellence across many fields. 
In ethics, this sits most naturally within the tradition of virtue ethics. The virtuous agent is, on the Aristotelian model, a person who sees and assesses situations rightly, and who responds to them -- in both action and feeling -- in appropriate ways: the virtuous person \textit{sees} the thing to do, does it, and feels good about doing it {\citep[\textit{Nicomachean Ethics}, pp. 1729--1867]{aristotle2009completev2}. Virtues traditionally associated with military practice include courage (both physical and moral), loyalty, honor, discipline, integrity, respect, and (of increasing prominence) virtues such as restraint, compassion, and mercy \citep{olsthoorn2011military,sparrow2013war,renic2018uav}. Virtue ethics \textit{can} be conceived as directly providing action-guiding rules: virtues as generating prescriptions, vices as generating prohibitions \citep[pp. 36--39]{hursthouse2010virtue}. Yet it need not be taken in this way. A more particularist approach (e.g., \citep{mcdowell1998mind}) may focus on the sensitivity of the virtuous person to uncodifiable but ethically salient factors in specific circumstances: the virtuous person sees what is called for, in a given situation, in a way that resists systematic formulation and in no way appeals to rules (indeed is often suspicious of the very concept of following a rule). This sort of view might equally be cashed out simply in terms of the virtue of \textit{practical wisdom}. Practical wisdom is understood in various ways, but can be broadly taken to be a meta-virtue allowing agents to negotiate issues not amenable to precision, codification, or rule-application -- e.g. identifying the ``mean'' between excess and deficiency in relation to a moral virtue (e.g., balancing bravery with recklessness), resolving tensions among conflicting recommended courses of action, or recognizing and weighing morally relevant features of a situation.\footnote{On practical wisdom, see \citep[\textit{Nicomachean Ethics}, Book VI]{aristotle2009completev2}; \citep{russell2009practical,miller2023three}.}

On this conception of virtues and virtuous agents, judging the right thing to do is not primarily an algorithmic process of deliberation. The virtuous person's good judgment is a matter of seeing situations rightly and responding appropriately, driven by good character, virtues, well-trained intuitions and experience. Rule-based processes may be necessary when it is not clear, even to an experienced agent, what is to be done. But, on the present view, such cases are the exception (for the virtuous person).

Rejecting a formulaic, rules-based approach to judgment and decision-making when dealing with opaque AI systems is important for at least three reasons. First, intuitive sensibility and practical judgment make possible insights and judgments that are not possible, we contend, for AI systems. A military targeting system may correctly identify enemy combatants as legitimate targets who \textit{can} be attacked, while missing morally salient factors raising questions over whether they \textit{should} be attacked. A system designed to identify targets cannot possibly be trained on \textit{every} factor that could influence a decision to deliver lethal force \citep{michel2020black}. Examples of this include so-called ``naked soldier'' cases, in which decisions whether to attack legitimate targets are muddied by unexpectedly mundane situational factors such as someone bathing or enjoying the sunshine \citep{walzer2006just,restrepo2019naked,restrepo2020defense}.\footnote{Cf. \citep{zajac2022spare}. \citep{feuer2024clothing} discusses naked soldier cases with respect to AI-enabled systems in the military. Note that our point here concerns AI systems' blindness to situational factors surrounding the legitimacy of killing naked soldiers, not whether such killing is indeed permissible (for such discussion, see \citep{zajac2022spare}).} Moreover, with opaque systems, it is uncertain what factors have been considered when arriving at outputs, undermining confidence in those outputs.

Second, whether or not AI systems are capable of identifying salient situational factors, humans are -- or can be trained to be -- \textit{extremely good} at identifying these. The Aristotelian tradition emphasizes that developing the virtues is not a matter of learning rules, but of habituating the right behavioral and emotional responses to situations (e.g., learning to take pleasure in virtuous actions and to feel ashamed of disgraceful ones \citep{burnyeat1980aristotle}). It is noteworthy here that apparent limitations can, in this respect, be strengths. For instance, the fact that human cognitive limitations necessitate reliance on heuristics, analogies, metaphors and the like, can support us in identifying unexpected similarities or analogies between apparently disparate situations. Humans are (or can be trained to be) excellent at identifying ``what matters'' in a situation, even if the concept of ``what matters'' cannot be fully articulated.\footnote{In the military, this is highly relevant for what is known as ``operational art''. See \citep{vego2017operational,vego2025operational,lyons2025decline,sookermany2026military}.} Humans can also be halted by something ``that matters'', even when not on the lookout for ``things that matter''. Sensitivity to what matters thus works in both directions.

Third, military personnel are already trained to develop situational awareness (the ability to rapidly and accurately size up a situation being a basic competence of good soldiering). 
Therefore, the fact that military personnel may be required to learn new awareness skills in order to work effectively with opaque and autonomous systems does not represent a significant disruption to how things are currently done.

We have, then, the vision of a person who, through judgment based on finely-honed intuition and experience, can size up situations and judge what is the thing to do, sensitive to ethical and other salient contextual factors. This can be characterized as the operationalization of virtue, virtuous perception, or practical wisdom; it is not the application of a rule, algorithm, or inferential move of the kind that could be programmed into a system. 
This is not to dismiss the importance of rules or technological knowledge and know-how, but it is to recognize that such forms of knowledge are limited.

In all practical fields, rules generally admit exceptions. Recognizing exceptional cases is not (or is not only) a matter of applying a further rule. Especially in the military, where rules abound, it is strong intuition, extensive experience, practical wisdom, and good judgment that makes right interpretation of the rules possible \citep{rebera2025virtues}. As AI systems are increasingly deployed, the need of these skills and forms of judgment motivates a renewed appeal to effective human-AI teaming, with each kind of agent playing up to the particular strengths in which they enjoy a comparative advantage.

What makes the human capacity for intuitive situation-sensitive insight and judgment so important is that it can be turned onto the opaque systems themselves. That is, an AI system can be taken to be either the object of human judgment or part of the context in which a human agent makes judgments and decisions. And if the problem with opaque systems is that they are not susceptible to direct interpretive analysis of their processes, it is at least possible that their performance can become the object of precisely the kind of wise, intuitive understanding and insight that humans specialize in. The system may not be transparent, but expert human judgment often involves the kind of intuitive understanding that transcends transparency. For instance, the distinction between an AI system being ``richly teamed with'', as opposed to merely ``used by'' a skilled operator is analogous, according to \citep{wood2024explainable},\footnote{\textit{Citation of author's unpublished manuscript removed for anonymity.}} to the way in which a skilled animal handler knows their animal -- being sensitive to various ``subtle factors and cues'' that alert them to what the system or animal is doing, is likely to do, why it did something, what its current status is, and so on \citep[p. 8]{wood2024explainable}. Training an operator to work with an AI system -- whether opaque or not -- is, on this model, partly a matter of ``getting to know'' it (as you might ``get to know'' a new colleague). 
Techniques such as XAI might support the \textit{getting to know you} process, but they cannot be all there is to it.

Part of getting to know a system is becoming familiar with the extent of its capacities, understanding what it can(not) do, and under what circumstances; learning to read the signs that signal that it is operating at the edge of its capacity or that it can take on more or greater loads. Generally, getting to know a system involves developing an accurate understanding of its capacities and of the conditions under which it can be relied upon to perform optimally.\footnote{\textit{Citation of author's unpublished manuscript removed for anonymity.}} Effectively and richly teaming with an AI system involves accurately understanding its capacities in contrast to, and in combination with, one's own (or those of one's team). It involves recognizing when and under what circumstances (and with what exceptions and limitations) human and machine are interchangeable. Here, the importance of not only traditional moral and military virtues becomes apparent, but also \textit{intellectual virtues}. These may include virtues such as algorithmic literacy, epistemic courage, or intellectual humility.\footnote{Following \citep{vallor2016technology}, we may think of these as, or in close relation to, ``technomoral virtues''.}

AI systems may be more \textit{reliable} than humans for certain tasks but, because unable to take \textit{responsibility}, less well suited to them overall. Alternatively, humans may clearly be the better choice for certain roles up until the point at which the workload outruns human limits of cognitive capacity and attention -- at which point AI systems become plausible replacements. In such cases, good judgment is needed to understand the capacities and limitations of both players. One must allocate tasks between humans and machines, recognizing that AI systems can replace humans only insofar as \textit{reliable task completion} is at stake, but never insofar as \textit{responsibility} is to be allocated. When a human role is occupied by AI systems, these replace only the functional aspects of that role. Responsibility may be redistributed, but it is never eliminated. This redistribution should be explicitly and thoughtfully managed -- yet another matter of sound judgment.

It is to be noted that the intelligence and judgment required to be able to manage this redistribution of responsibility effectively is itself situation-specific. In particular, the profile of a particular role -- and hence the possibilities for AI systems to contribute to it -- will appear differently depending on one's position in an institutional hierarchy. In the military, the decision to deploy opaque AI systems might be taken at senior commander levels, based only on generic understandings of the capacities of the systems and of the units into which they will be deployed, whereas unit-level deployment decisions may be taken with an eye to which specific team members best know the systems, which are best equipped to deal with them under expected circumstances, and so forth. The distinctively human capacity for making moral and other evaluative judgments based on an uncodifiable sensitivity to the particularities of given situations is precisely what is required to ensure that the introduction of opaque systems -- which are powerful but lack this capacity -- is sensitively handled. And this is made possible by turning that uncodifiable sensitivity and judgment into an asset for managing deployment of opaque systems.

\section{A Humanistic Response to Opacity's Challenges}\label{sec_humanism} 

Practical wisdom, practical judgment, and sensitivity are essential tools which complement technical and rules-based approaches to responsibly deploying opaque and autonomous systems. And for any such deployment, the first and most important point where judgment is necessary is within the first basic decision: Do I deploy this system? Taking for granted all of the risks and challenges of opacity sketched in Section \ref{sec_opacity}, all other things being equal, the default response on ethical, legal, and prudential grounds should be \textit{not to deploy} an opaque system. Thus, unless the opaque system clearly improves outcomes, streamlines processes, or reduces the necessary inputs for a given output, deployers should opt for a deterministic system or direct human action. But what about when the deployment of an opaque system is expected to improve the situation in some fashion? How can wisdom, judgment, and sensitivity be beneficial for mitigating the risks of opacity? To answer these questions, we must recognize and unpack a critical aspect of judgment, namely that \textit{judgment is a resource}. 


\subsection{Judgment as a Resource}

Within any given window of time, a particular individual will only have so much capacity for judgment which they can use. Some, by nature or by training, may have more judgment, or be able to replenish their judgment more quickly. Individuals may also require different amounts of judgment for distinct tasks; e.g., a veteran commander may be able to quickly, reliably, and with little effort make sound judgments based on intuition and the information available, while more novice commanders may expend much judgment in the same scenario, simply because they lack the experience and intuitions to decide easily, experience and intuitions which, respectively, feed into and come out of practical judgment. The amount of judgment one has at any given moment may potentially be depleted -- due to fatigue, cognitive overload, judgment impairment through emotional stress, etc. -- at which point decisions will become confused, information will be overlooked, and actions which are potentially dangerous may be taken. When judgment is depleted, it is necessary that decision-makers reduce how much judgment they must use -- by reducing their cognitive and decisional load -- or they may be required to fully withdraw from a given task and replenish their stocks of judgment -- bluntly put, a nap is sometimes the operationally sound decision. Knowing how much judgment one has, how quickly it replenishes, and when it is responsible to offload tasks in order to marshal one's reserve of judgment are all things requiring not just technical or institutional fixes, but which demand individuals have well-developed virtuous character, understand themselves and their missions well enough to know where their limits lay, and possess the courage to place appropriate limits on their own decision-making. 

Taking these considerations and applying them to the question of whether to deploy opaque systems at all, we arrive at a contextually-grounded conclusion: users should consider deploying opaque systems when they assess that, were they not to, they would deplete their pool of judgment needed for responsibly executing current tasks and they would not be able to replenish (enough of) it as is needed for upcoming tasks. For example, suppose a military commander is faced with a developing conflict scenario where an adversary has not made any declarations of hostilities and has not yet struck the commander's forces or home territory, but manned and unmanned aerial and ground units are being mobilized at scale. The commander feels competent to guide defensive efforts in response to the current threat picture she is facing -- she deems her pool of judgment and its replenishment rate to be sufficient for making assessments over all critical decisions she must make. However, she also assesses that if the enemy adds another type of threat, opens up another front for action, or begins sabotage actions in her rear areas, then she will no longer be able to reliably and safely exercise the required level of judgment in addressing the full array of threats she is facing. Assume also that the deterministic systems at her disposal are not able to sufficiently reduce the  overall amount of judgment she has need of, and that all her subordinates and superiors are in similar situations and do not have excess capacity which she might leverage (i.e., the deployment of an opaque system represents the only viable alternative for retaining control).

In such a case, the commander could continue as she currently is, using her own judgment to make decisions about tactical and operational responses and not deploying or relying on any opaque systems. In the event that the enemy does carry out actions which overload her capacity to responsibly exercise judgment over all her areas of responsibility, she will however have to decide what to do. She could still continue as is, not deploying opaque systems, and as a result seeing a decline in ability to effectively and responsibly judge what to do in response to the threat picture. The decline in ability may not be immediately apparent, as small lapses may not be noticed, but the cumulative effects of fatigue and cognitive overload may create compounding failures leading to significant mistakes and loss of life. Alternatively, she could offload portions of analysis or decision-making to opaque systems. Such off-loading may also be undertaken in different manners. For example, she could deploy systems which automate some portion of the analysis of intelligence, surveillance, and reconnaissance (ISR) information being acquired for all objects and potential targets in her theater of operations. Alternatively, she could deploy narrower opaque systems which only overtake specific tasks, e.g., launch interceptors autonomously when pre-set triggers are met. She might also delegate the full targeting cycle to an opaque and fully autonomous system, but do this for only a limited set of targets, e.g., incoming missiles and unmanned aerial vehicles. 

In any event, the commander contemplating the deployment of opaque or autonomous systems or the delegation of critical tasks to them will be presented with an array of potential deployment options and parameters. And the choices over each of these will impact on how much of her judgment she might save for other tasks going forward, and how much she must use now in deciding what system(s) to deploy and under what parameters or limitations. Given the importance of speed in making those higher-level judgments, it is critical the commander has a well-developed character allowing her to confidently and courageously recognize her own limits and judge with those in mind. Because her judgment in that moment is most centrally anchored \textit{in the assessment she gives of herself}, it is also not something which is amenable to general rules an organization might develop for all. 
Rather, the character and virtues of the commander herself will be central to this highly subjective assessment, an assessment which crucially hinges on humanistic values such as self-reflection, honesty, integrity, courage (in admitting one's limits), and judgment. None of these are machine-translatable or fully operationalizable through technical ``solutions'' such as explainability, and institutionally fixed rules cannot guarantee responsible deployment. That being said, technical features of a system (such as explainability or interpretability) can help commanders better assess what parameters or limits on deployment may be needed, both in general and in response to commanders' own limits and needs, and institutional rules can provide general guidance and information serving as checks on decision-makers' intuitions and snap judgments (when intuitions greatly diverge from institutionally set rules, decision-makers are given impetus to reconsider their thinking, as critical factors may have been overlooked). 


Rules and system design will serve as core enablers and multipliers of how well judgment can be leveraged to improve outcomes, but at the end of the day, rules and design cannot by themselves underpin responsible and effective use of any system, opaque or otherwise. Rather, judgment will always be necessary. And when systems are opaque, autonomous, or both, judgment provides a means for some of opacity's challenges to be mitigated. Such mitigation is further improved when technical and institutional approaches are also present, but the central need is always judgment, which in turn is developed and entrenched through character and virtues. 

\section{Human Limitations and Deployment Limitations}\label{sec_limits} 

Focusing on humanistic traits and values such as character, virtue, judgment, or intuition indicates not just positive approaches to responsibly deploying opaque systems, but also limits to be taken into consideration. In particular, human limitations must be taken seriously and translated into deployment limitations. Though by no means an exhaustive accounting, there are three central human limits worth briefly discussing. 

First, virtually no user of an opaque AI-enabled system will be fully knowledgeable about the technical aspects of the system and also knowledgeable in their domain of expertise and application; e.g., a commander may be highly knowledgeable concerning military science but only minimally technologically literate. 
Some technical knowledge will always be achievable for professionals of various stripes, but we should not expect any individuals using opaque systems to be as knowledgeable about them as developers would be. 

Limited technical literacy does not, by itself, constitute a fundamental challenge to deploying systems. However, the degree of individual users' technical (il)literacy should shape their readiness to rely on systems and the precautions they deem necessary when deploying them. Thus, a commander who has well-cultivated practical judgment and experience in their command role, but who knows they are very uncertain about how a particular opaque system functions, should default to not deploying that system unless they believe failure to do so will lead them to rapidly depleting their well of judgment and then making dangerous -- and potentially preventable -- mistakes. That being said, even in that case, the commander should try to find other deterministic systems to rely on first, delegate tasks to human subordinates, or try to rapidly gain additional technical knowledge to guide their deployment and use of the opaque system. The latter point is worth highlighting, as technical literacy is not just important for basic decisions about whether to use opaque systems at all, but also with regards to what parameters and limitations on deployment a user may find appropriate. 

The opacity of some systems, especially those with autonomous functionality, can make them unpredictable in their outputs, and limits on deployment represent a critical way to mitigate that risk. But setting those limits requires minimal understanding of how the system functions, what mistakes it might make, with what severity, etc. If a potential deployer lacks that understanding, then they ought not deploy the system. If, however, they have some understanding, but not a great deal, then their deployment, and the limitations governing it, should reflect this. I.e., geo-temporal limits, conservative decision protocols, and the like can serve to buttress uncertainties about system functionality. These should still though be relied on only as a last resort when the user deems that they must deploy that opaque system because all other options are worse and would lead to degradation of their ability to competently and responsibly judge in critical decisions going forward. 

A second limitation users may suffer from are limits in their own individual judgment or limits which make decisions more costly for their pool of judgment, reducing overall judgment available. These may be short-term and temporary (fatigue or context-dependent anger may limit judgment), medium-term and open to amelioration (lack of training or experience may make decisions more ``judgment-costly'', reducing overall judgment available), or longer-term and more difficult to address (weaknesses in basic character traits or virtues -- such as self-reflection, honesty, or courage -- can both limit judgment in total and make individual decisions more judgment-costly). 

If potential users of opaque systems are lacking in basic virtues and character traits necessary for effectively gauging their own capabilities and limits, they may need to be fully removed from decision-making over the deployment of opaque systems (see Section \ref{sec_deployment} below). This is because, due to their lack of self-reflection, honesty, courage, etc., they are not able to adequately or reliably determine what systems to deploy, when, or under what limits/parameters. Simply put, practical judgment is necessary for responsible use of opaque systems, as it is a prerequisite for being able to make use of technical tools (like explanations) or implement institutionalized rules (like law or doctrine). And if a commander cannot do these things, then they ought not have at their disposal systems whose use demands that. I.e., the system should, at the level above the potential deployer, be limited so as to make it unavailable to the limited individual. 

For less severe and long-term limits, like temporary fatigue or present (but not enduring) lack of experience, deployers should, if possible, delay the use of opaque systems until they have recovered enough judgment to decide well or have acquired enough experience to develop the instincts, intuitions, and practical wisdom necessary for judging how and under what limits to deploy the system. In these cases, the limits to deployment may be merely temporal, not indicating that the system should not be deployed at all or that it must always have some limits, but rather that the user should delay deployment until their own temporary limits have been addressed.

The final human limit to consider is the range of cognitive biases we possess and which impact on our use and reliance on AI-enabled systems, algorithmic tools, and even simply machines as such.\footnote{Our concern is human biases, but there are biases within AI as well which impact responsible deployment. See \citep{nelson2019bias,roselli2019managing,gichoya2023ai,zhou2024bias,nudo2026generative}.} 
Biases or cognitive limits centrally relevant to this are action bias \citep{patt2000action,jeremiah2025making}, automation bias \citep{cummings2017automation}, confirmation bias \citep{nickerson1998confirmation}, or framing effects \citep{tversky1981framing,nelson1997toward,druckman2001evaluating,druckman2001limits}, though this is by no means an exhaustive list.

Each bias we (may) have can indicate for different limits to deployment, and even differences in where and when these limits should be set. For example, framing effects and confirmation bias will be highly relevant during the use of AI-DSS \citep{arnott2006cognitive,phillipswren2019cognitive,minotra2024reviewing,talbert2026triage}, but may have no effect on the deployment of an autonomous system for, say, the full targeting cycle of incoming missiles and unmanned aerial systems. However, automation bias may impact on how we view the risks of offloading a targeting cycle to an opaque autonomous system, potentially tainting our assessments of necessary limits. Action bias may also lead us to more favorably view systems which produce clear and striking outcomes when compared against those which do not, even though the latter may be operationally and strategically superior. 

Given that each bias raises distinct challenges, and that each of these will interact in different ways with different systems, and also with variance due to different missions or tasks, it is impossible to explore these fully here. Suffice to say though that these myriad complicating factors highlight the importance of practical judgment, practical wisdom, and a virtuous character underpinning decision-making under stress. When time is short and lives are on the line, it may not be possible for an unwieldy manual listing biases, systems, and deployment contexts to be consulted to determine what deployment limits might be needed. But a commander with a well-developed character, practical wisdom, and honest self-reflection may be able to quickly determine a minimally responsible course of action which retains effectiveness in deployment. 

\section{Objections and Deployment Considerations}\label{sec_deployment} 

For virtue and practical judgment to usefully mitigate the challenges of opacity, approaches to developing these and ensuring their robustness in high-stakes and high-stress decision-making contexts are necessary. Virtue ethics and humanistic approaches are already present in society at large and in critical domains where life-and-death decisions must be made (e.g., in the military \citep{olsthoorn2011military,robinson2016introduction,robinson2016ethics,schulzke2016rethinking}, medicine \citep{seoane2016virtues,kotzee2017virtue,lyon2021virtue,doukas2022virtue,kim2024professional}, or disaster response \citep{webb2004role,mendonca2006training,morton2007great}). Given this existing body of work, we will set aside basic elements of virtue ethics education and instead focus here on two lingering objections to our proposal and the deployment considerations they point toward.

Throughout, we have stressed the importance of virtues and practical judgment in the deployment of opaque systems, but above we noted that not all individuals may have the necessary virtues/judgment to do this responsibly or effectively. We stated that such individuals may need to be fully removed from decision-making over the deployment of opaque systems, but a critic might wonder what that means in practice. To address this, it is worth noting the hierarchical structures of most critical domains, especially the military, which is our main focus. In this area, systems are deployed at different levels, by different individuals, with different skill sets, virtues, and levels of judgment. Additionally, not only are individuals higher in the hierarchy expected to have more experience and judgment surrounding their domain and the tools they may use, but also with regards to their subordinates. And tools might also be stratum-dependent, with the authority to deploy (opaque) systems tied to particular strata of the domain. 

Concretely, in the military there are strategic, theater-level, operational, and tactical decisions which might be made, and some tools or systems are tied to one or more distinct levels. Decisions about the use of opaque weapon systems may be tactically and operationally relevant, but not something theater-level commanders are as concerned with. Moreover, theater-level commanders may not be able to be concerned with such decisions, as this would be another decision drawing on their pool of judgment (which is finite). Because of this, higher-level commanders delegate this to subordinates more closely tied to the discrete deployments of such systems. Those subordinates, due to proximity of deployment, are also better situated to exercise practical judgment about deployments of that type of system. 

Importantly, every higher level of command is expected to have a sufficient understanding of their subordinates to know what strengths they have as well. Some lower commanders may be deemed to have enough virtue and practical judgment to be entrusted with decision-making over the deployment of some opaque system, while others might be given explicit orders to (not) deploy the system to prevent mistakes being made at that lower level. When an individual lacks the virtue and judgment to make such decisions themselves, a superior in the institutional hierarchy should be taking that decision away from the subordinate. This is also nothing new, as superiors are always expected to extend and withdraw latitude in keeping with both longer-term and temporary limitations subordinates may suffer. For example, a commander who recognizes a combatant is in distress should give that combatant explicit orders about what they may (not) do, in order to lower the likelihood of their distress leading to mistakes or unwanted outcomes. At a meta-level, the decision space about whether and how to deploy opaque systems should thus be tailored to the subordinate having that decision delegated to them, with superiors leveraging knowledge of their team to ensure individuals unfit to make certain decisions are not given power to do so. 

A second and central concern in the deployment of any AI-enabled system is how this may affect the skills and capabilities of user/deployers over time. I.e., does use or reliance on the system lead to de-skilling? 
In response, we would highlight that our proposal is not that some system be relied on, but, on the contrary, that one defaults to not using a system unless the situation leads one to believe judgment will be depleted before all necessary or critical choices have been made. Thus, in keeping with a virtuous epistemic stance and in line with one's practical judgment, the first and most basic consideration is whether delegating some task to an opaque system is necessary or sensible at all. Further, one is enjoined to only deploy if practical judgment leads one to believe judgment will deplete unless decisions are delegated. Following this stance, a corollary is that whenever a task has been delegated, the delegating individual should, at the end of the timeframe of decision-making, be very near to having no more capacity to reliably make judgments. Illustratively, if we consider individuals as having a ``judgment meter'' ranging from 100 (full capacity for judgment) to 0 (no capacity for judgment), if some tasks have been delegated to an opaque system, then at the end of the mission or day, the individual who did that delegation should be very near to having 0 judgment left. If that is not the case, then they delegated unnecessarily, and in keeping with a virtuous character, they should reflect on how to better assess situations and themselves in the future, with an eye to not delegating prematurely or unnecessarily. Such a focus on first decisions of whether to deploy systems at all combats de-skilling concerns by making use of systems less regularized and anchoring the importance of judgment at each decision point. 

A further and more insidious de-skilling concern is whether a machine might always take over a single type of task, thus de-skilling one narrow skill that is routinely the first thing to be delegated away. System design and capability may feed into this worry, with some particular system(s) always taking priority in the delegation pipeline due to those systems, e.g., freeing up the most judgment for the decision-maker trying to navigate complex and fraught domains. Unfortunately, there is no clear and surefire method for combating this concern, partly because of the fact that a system which frees up much judgment capacity should in many respects be given priority by decision-makers. Focusing on virtue, character, and self-reflection may help to mitigate this worry though by enjoining individuals to consider how their own capacities may be being degraded over time. Thus, a focus on humanistic values like virtue and judgment may alter de-skilling (of technical skills) to a scenario of re-skilling (to humanistic skills)\citep{vallor2013future,vallor2014deskilling}.

\section{Conclusion}\label{sec_conclusion} 

Opaque AI-enabled systems, especially those embedded in autonomous systems, raise numerous and deep worries regarding their responsible and effective deployment. We have argued that virtue, judgment, and intuition may be leveraged to combat many of these challenges. In closing, it is worth stressing that our proposal is not intended as an alternative or competitor to technical and institutional approaches. Rather, humanistic values are complementary to existing frameworks, and indeed necessary for those frameworks to fully bear fruit. Moreover, focusing on the traits and strengths of humans showcases an often overlooked set of values which can be crucial in mitigating worries associated with emerging technologies. Our humanistic approach should thus be viewed through the ``Swiss cheese'' safety model found in engineering disciplines, where each layer of safety is not meant to fully eliminate risks, but rather to shift where the dangers may lay, so that no set of gaps line up fully to allow risks to materialize \citep{larouzee2020good,wiegmann2021understanding}. Humanistic values will not solve the issue, but they shift the space of dangers, and when coupled with technical and institutional approaches, may allow for responsible and effective use of opaque systems in critical domains. 

\pagebreak 

{\small \bibliography{master}}

\end{document}